# Electrical transport and optical studies of ferromagnetic Cobalt doped ZnO nanoparticles exhibiting a metal-insulator transition

M. Naeem, *S. K. Hasanain, A. Mumtaz.

*Magnetism and Superconductivity Group, Department of Physics, Quaid-i-Azam University*

*Islamabad, Pakistan.*

**Abatract:**

The observed correlation of oxygen vacancies and room temperature ferromagnetic ordering in Co doped $ZnO_{1-\delta}$ nanoparticles reported earlier (Naeem *et al* Nanotechnology **17**, 2675-2680) has been further explored by transport and optical measurements. In these particles room temperature ferromagnetic ordering had been observed to occur only after annealing in forming gas. In the current work the optical properties have been studied by diffuse reflection spectroscopy in the UV-Vis region and the band gap of the Co doped compositions has been found to decrease with Co addition. Reflections minima are observed at the energies characteristic of $Co^{+2}$ d-d (tethrahedral symmetry) crystal field transitions, further establishing the presence of Co in substitutional sites. Electrical transport measurements on palletized samples of the nanoparticles show that the effect of a forming gas is to strongly decrease the resistivity with increasing Co concentration. For the air annealed and non-ferromagnetic samples the variation in the resistivity as a function of Co content are opposite to those observed in the particles prepared in forming gas. The ferromagnetic samples exhibit an apparent change from insulator to metal with increasing temperatures for T>380K and this change becomes more pronounced with increasing Co content. The magnetic and resistive behaviors are correlated by considering the model by Calderon *et al* [M. J. Calderon and S. D. Sarma, Annals of Physics 2007 (Accepted doi: 10.1016/j.aop.2007.01.010] where the ferromagnetism changes from being mediated by polarons in the low temperature insulating region to being mediated by the carriers released from the weakly bound states in the higher temperature metallic region.

* skhasanain@qau.edu.pk

## I. INTRODUCTION

The continuous attempts to develop next generation devices equipped with multifunctions are now being extended to the search for materials that can combine magnetic, electronic, and photonic responses. One example of such efforts is the quest for a ferromagnetic material that can inject spin-polarized carriers into semiconductors. Ferromagnetic metals and alloys, such as Fe and FeNi, have been found to be inadequate, since spin-polarized carrier injection was found to be difficult due to resistance mismatch [1]. In attempts to overcome such problems, dilute magnetic semiconductors DMS, such as Ga(Mn)As, have emerged as good candidate materials [2] Currently numerous materials, including Cd(Mn)GeP [2,4] Ga(Mn)N, [5] Zn(Co)O [6] and Ga(Mn)P [7] have been reported to exhibit room temperature ferromagnetism (RTFM). In spite of extensive efforts in this area, there has been a great deal of controversy, especially on fundamental issues such as the origins and characteristics of the observed FM [8,9]. However there is considerable evidence to the effect that oxygen related defects play a significant role in stabilizing room temperature ferromagnetism (RTFM). There is also some experimental evidence associating the presence of oxygen vacancies ($V_o$) with the existence of FM in TM doped ZnO. We have shown earlier that Co-doped ZnO nanoparticles exhibited RTFM by annealing in reducing atmospheric pressure, which was attributed to oxygen vacancies [10]. M.Venkatesan *et al* [11] have reported a correlation between the magnitude of magnetic moments and the oxygen partial pressure during annealing, with higher partial pressure reducing the amount of magnetization. In addition, Kittilsved *et al* [12] also report the observation of RTFM in Co doped ZnO nanocrystals with oxygen containing surfaces and proposed that carriers generated by oxygen vacancies mediate the magnetic interaction. Rode *et al* [13] prepared films of $Zn_{0.75}Co_{0.25}O$ in low oxygen partial pressure ($10^{-6}$ Torr) that were found to be ferromagnetic at room temperature. Recently Coey *et al* [11,14] find a connection between the electronic structure of ZnO based DMS and its ferromagnetic properties, which is explained in terms of the spin-split donor impurity-band model. The likely origin of the donor impurity band in ZnO is lattice defects, such as oxygen vacancies, which can have trapped between one and two electrons ($F^0$ centers) [15-17]. While, it is clear that the extrinsic as well as intrinsic donors are known to affect the carrier concentration and FM [18], the exact nature and interdependence of these entities is far from understood. In this work, we report the optical and transport properties of highly insulating Co doped ZnO nanoparticles and their dependence on magnetic dopant concentration and forming gas annealing, explained in terms of the role of oxygen vacancies in this system. In order to understand the fundamentals of carrier generation and transport characteristics of high temperature ferromagnetism, we have investigated the temperature dependence of the resistivity of Co doped ZnO nanoparticles with varying Co concentrations. An insulator to metal transition (IMT) is observed at higher temperatures and its origin is discussed in terms of a possible degenerate band formation due to the oxygen defects.

## II. EXPERIMENT

Nanoparticles of $Zn_{1-x}Co_xO$ (x=0.00 to x=0.06) were synthesized by heating metal acetates in organic solvent following the previously procedure reported [10]. For each concentration of Co, the samples were annealed in air and in the forming gas (Ar95%+H5%), separately. The samples were characterized by transmission electron microscopy (TEM), and UV-Vis diffuse reflectance spectroscopy (DRS). Supportive results of x-ray photoemission spectroscopy (XPS) and XRD, have been published earlier [10]. Here we recall that no secondary phase or metal related peak were detected within the sensitivity of XRD. However, lattice d(002) spacing were observed to increase linearly. The slightly increasing trend in lattice with increasing 'Co' concentration revealed that the ionic size of tetrahedrally coordinated $Co^{+2}$ is larger than that of $Zn^{+2}$ in the same symmetry as discussed in other works [19, 20]. Alongside the structure analysis, the XPS measurements [10] also revealed that within the solubility limit of our preparation route (x = 0.06), $Co^{+2}$ ions are tetrahedrally coordinated with oxygen in the ZnO lattice.

Bright field transmission electron micrographs (TEM) of both samples annealed in forming gas and air are shown in figure 1a and 1b respectively. Both samples show that the particles are generally round and faceted with typical particle size ranging between 20-50nm but larger size particles are also visible. No differences in size were observed by annealing in different atmosphere (air and forming gas). Keeping in view the very large resistivities of the samples, typically of the order of MΩ-kΩ-cm, the two-probe method was employed to study DC resistivity of palletized nanoparticles using a self made setup. The magnetic properties of both types of samples, those annealed in air and in forming gas, have been reported earlier [10]. Only those samples exhibited ferromagnetism which had been annealed in a forming gas and show clear hysteresis loops with coercivity and remanence at 300 K. However, no evidence of ferromagnetism was found in air annealed samples down to 10K, the lowest temperature investigated. All samples annealed in forming gas exhibited RTFM with the moment (emu/g) monotonically increasing with increasing concentration of cobalt. The maximum moment per Co atoms was $\mu \sim 0.25\mu_B$/Co (4% Co doped) at room temperature. The value of magnetization ranges from 0.141(x=0.02) to 0.25$\mu_B$/Co (x=0.04). These values are comparable with some of the typically reported ones in the literature [21]. The coercive fields *(*$H_c$*)* of Co-doped samples were found to be typically ~320 Oe at room temperature. The relatively large coercivity values we observed were explained [10] on the basis of single domain nature of these nanoparticles leading to the magnetization rotation magnetic mechanism [22]. Larger sized particles (55nm) of the same composition showed coercivities typical of DMS in the range of 30-50Oe [23].

### III. RESULTS AND DISCUSSION

Optical characterization of $Zn_{1-x}Co_xO$ (0.00 ≤ x ≤ 0.06) nanoparticles were carried out by measuring the diffuse reflectance spectroscopy (DRS). All spectra were taken in the range of 250-1000nm on Lamba-950 Perkin-Elmer spectrophotometer with integrating sphere attachment and Spectralon reflectance standards. Figure 2a shows, in addition to the reflection minimum corresponding to the band gap, three well defined reflectance minima at 660, 615, and 568 nm respectively. These minima are in agreement with the $Co^{+2}$ d-d (tetrahedral symmetry) crystal field transitions 4A2(F) → 2A1(G), 4A2(F) → 4T1(P) and 4A2(F) → 2E1(G) respectively [6, 24] where A, E and T are generally designations of intermediate energy bands [24]. The appearance of these transitions and their growing sharpness with increasing Co content supports the inference that the doped $Co^{2+}$ ions are in the high spin (*S*=3/2) state in a tetrahedral crystal field symmetry [24]. We further observe that with increasing Co content there is a trend of decreasing reflectance (increasing absorption) in the visible region of the spectrum consistent with higher absorption by the Co ions due to the aforementioned transitions in tetrahedral symmetry.

For analysis purposes the diffuse reflectance, *R*, of the sample can be related to the Kubelka-Munk function *F(R)* by the relation $F(R) = (1-R)^2/2R$ [25]. The energy band gap of the $Zn_{1-x}Co_xO$ nanoparticles were determined from the diffuse-reflectance spectra by plotting the square of the Kubelka-Munk function $F(R)^2$ vs. energy and extrapolating the linear part of the curve to $F(R)^2$=0 as shown in figure 2b. This yields the direct band gap energy. The optical band gap of the undoped samples was determined by the above method to be 3.34±0.01eV. The absorption edges are seen to be shifted towards higher wavelengths/lower energies with increasing Co content as shown in the figure 2a. This would clearly suggest, as shown in the figure 2c, a lowering of the band gap with increasing Co content upto a maximum decrease of 0.21eV for 6% Co. Such compositional dependant shift in the band gap energy has been recently reported in Co doped ZnO nanorods [26]. This change in the band gap has been explained on the basis of the variation of the lattice parameters due to the effects of doping. It has been observed that the energy band gap decreases with increase in the lattice parameter [27]. This explanation is consistent with our observation [10] that the lattice constant d(0002) *increases* with increasing Co concentration. Therefore the observed decrease in the energy gap with the increase of Co concentration can be explained on the basis of increase in lattice parameter due to increase in Co concentration. The variation of the band gap (red shift of band gap) with increasing Co content has also been related in the literature to the increased sp-d exchange interactions between the band electrons and the localized d electrons of $Co^{+2}$ cations [28]. In addition to the variation in the band gap with Co doping a small but reproducible step was observed in the reflectance spectra of all the Co doped samples at λ~ 400nm (3.1eV) close to the band edge (figure 2a). This feature is shown expanded in the inset of Fig.2a. It is noticeable that the step feature becomes more pronounced with increasing Co content. Similar features in the optical spectra of transition metal doped ZnO have been reported in the literature and they have been related to the development of dopant related defect levels near the band edge [26,29] that are responsible for the increased absorption that is manifested as a step in the reflectance. In our case the presence of this step suggests that the addition of Co in ZnO leads to the formation of defect levels close to the conduction band edge.

Resistivity (ρ) measurements were performed over the temperature range 300 to 480K to observe both the effects of the Co dopants and the forming gas on the electrical transport. Pelletized samples of the air and forming gas annealed were used for these measurements after further annealing in Ar at 300 $^0$C. The samples were compacted uniaxially into pellets under identical conditions so as to retain the same density over the entire composition range thereby minimizing the possible variations in the inter-grain resistivity contributions. While the resistivity studies were performed on the full composition range 0.00≤x≤0.10, only the data for the range 0.00≤x≤0.06 are being shown here. The resistive behavior of the omitted compositions (x=0.08 and 0.10) displayed marked differences compared to the others. In particular a strong metallic behavior was observed in these compositions consistent with the presence of the metallic clusters, as deduced from the XPS. Hence the metallic behavior in these omitted compositions is understood to be unrelated to an intrinsic source as opposed to the lower cobalt compositions to be discussed below.

Figures 3a and 3b represent the temperature variation of resistivity of samples that were annealed in air and in a forming gas, respectively. The data are plotted against 1/T and the fit to the Arrehnius equation, $\rho=\rho_o \exp(E_a/k_BT)$, is also shown for each case. We observed a marked change in the resistivity for the samples annealed in the forming gas as compared to those annealed in air. For the air annealed samples, both the undoped and all the Co doped samples are semiconducting over the entire temperature range measured, but it is apparent that the overall resistivity increases with increasing Co content. Room temperature resistivity varied from $26 \times 10^6 \Omega$cm to $108 \times 10^6 \Omega$cm for Co concentration varying from x = 0.00 to x = 0.06 as shown in figure 3a. The variations in the values of $\rho_o$ and $E_a$ obtained from the fitted lines in figure 3a and 3b are shown in figures 4 and 5 respectively. In the air annealed samples the prefactor $\rho_o$ clearly increases with increasing Co content consistent with the role of Co as a defect scattering center in the ZnO lattice. The activation energy $E_a$, derived from the fittings yields $E_a$=0.321-0.353±0.009 eV, also increases monotonically with increasing Co concentration as shown in figure 5. The overall trend of increasing resistivity with increased doping of a transition metal (TM) in ZnO is expected in general and is consistent with the literature [6,30] where it has been attributed to the increased defect scattering due to the addition of Co ions. The increase in the values of the resistivity prefactor $\rho_o$ (figure 4) with increase in cobalt concentration in the air annealed samples clearly reflects this effect. The increasing values of $E_a$ in the air annealed samples may be seen in the same spirit as a consequence of the enhanced defect scattering and the consequent requirement of higher energies for activation into conducting states.

The resistivity measurements of the samples annealed in forming gas however show major differences compared to their air annealed/non-ferromagnetic counterparts. The differences between the air and forming gas annealed samples can be seen in figures 3, 4 and 5. The temperature dependence of the resistivity of these samples (figure 3b) will be discussed later. Firstly we note that after annealing in a forming gas there is a remarkable decrease in the resistivity as compared to the samples annealed in air. This effect was seen to be clearly reversible i.e. on exposures of these same samples to flowing oxygen at $600^oC$, the original high resistance values were retrieved. This effect is shown in figure 6. We understand these reversible changes in the resistivity due to the reduction-oxidation, in both the undoped and Co doped ZnO nanoparticles, on the basis of our own [10] and other reported works [31] that annealing in forming gas leads to increasing oxygen vacancies which act as n-type dopants. The increased carrier concentration due to these defects leads to a lowering of the resistivity. Annealing in an oxygenating atmosphere removes the vacancies and the associated carriers and restores the high resistivity of the materials.

Secondly we note a variation of the resistivity with Co concentration (for samples annealed in forming gas) as shown in figure 3b and in figure 4 for the variation of $\rho_o$ for different concentrations at T=300K. In contrast to the observed variation with *x* for the air annealed samples, here we observed a systematic *reduction* in the overall resistivity and in the values of $\rho_o$ and $E_a$ with increasing Co concentration. The activation energies are seen to decrease from 0.309 to 0.270±0.004 eV over the given composition range as given in figure 5. Room temperature resistivity values varied from 3.0 x$10^6$Ωcm (x=0.00) to 0.2 x$10^6$Ωcm for x=0.06 respectively, while $\rho_o$ varied between 25 x$10^6$Ωcm and 5 x$10^6$Ωcm. If the overall magnitude of the resistivity is controlled, even partially, by the concentration of carriers contributed by the oxygen vacancies, the decrease of the resistivity with increasing Co content (for compositions annealed for the same time in the forming gas) suggests that a larger number of oxygen vacancies may be present in the higher Co doped samples. This would suggest that *Co dopants in ZnO promote the stabilization of oxygen vacancies which in turn help to decreases the resistivity*. Similar results have also been observed by J. Alaria *et al* [32] in Co doped ZnO nanopowder where it is shown that the concentration of oxygen related defects increases with increasing of Co dopant. This is explained with reference to the observation that the bonding of oxygen with the Co ions in tetrahedral symmetry is weaker than that of oxygen with Zn ions in ZnO. Furthermore the Co dopant (even at substitutional site) makes the lattice distorted, as evident in our XRD data, and hence the bonding strength of oxygen becomes weaker near Co ions as compared to the host ions (Zn) in ZnO lattice. This may result in enhancing the vacancy concentration with Co dopants. It has also been shown by simulations in a similar system that oxygen vacancies prefer to reside closer to the Co ions than the host metallic ions [33]. While the decrease in $\rho_o$ with increasing Co content in the samples (annealed in forming gas) may be reflective of the increase in the number of carriers, the simultaneous decrease of the activation energies could be indicative of the increased itinerancy of the carriers [34] generated by the vacancies. i.e. the vacancies generate more carriers and these latter are relatively weakly bound to the defects (oxygen vacancies).

An intriguing feature of the temperature dependence of the resistivity in the doped and forming gas annealed samples is the change from insulating to metallic behavior for T>380K, as shown in figure 3b. It can be seen that the extent of the metallic behavior increases very systematically with increasing Co content where it starts off initially as simply a decrease of the high temperature resistivity slope for 2% Co sample and then develops into a resistivity minimum of more pronounced sharpness as the Co content increases. This insulator to metal transition was not observed in the pure ZnO (annealed in air or forming gas) nor in any of the Co doped, air annealed samples. Furthermore the transition was seen to be reversible with oxygen removal and after annealing in an oxygen atmosphere (shown in figure 6) the transition was clearly absent. The metallic trend is thus clearly a combined effect of the increased carrier concentration and the association of these carriers with the Co ions. In terms of energy this temperature (380K) corresponds to about 32 meV, which is in the typical shallow donor range. Similar observations of high conductivities and positive temperature coefficient of resistivity (TCR) have also been reported in the case of Al:ZnO [35,36] , Nb:TiO$_2$, [37] and Ni doped ZnO [38]. Can the metallic behavior be a consequence of metallic clusters of Co? It is apparent that if at all there are any metallic clusters present in the system then these are below the resolution limit of the XPS. Given our beam size (1 mm) and the escape depth (10 nm), it is estimated that the system is capable of a detection limit of 500 ppm [10]. However to exhibit an overall metallic trend the cluster density would be expected to exceed the percolation threshold. It does not appear possible that a cluster density lower than the XPS resolution limit could still be above the percolation threshold in this system. Hence we rule out the metallic behavior as originating in percolating Co clusters. A similar and possible scenario to explain the change from an overall semiconductor to metallic behavior is to consider the resistivity as composed of an intergrain (semiconductor) and an intragrain (metallic) part. While this picture can not be ruled out altogether, it appears to us to be unlikely given the systematic of the changes with Co content and forming gas annealing. We note that the change from an overall semiconductor to metallic behavior continues to occur around the same temperature (T~ 380K) despite increasing Co content. This suggests that this transition temperature is related intimately to the generation of free carriers (bound to shallow doping states). For the trend to be due to an overlapping of grain and intragrain parts it is very unlikely that the transition point, where the two contributions intersect, would not shift in temperature with changes in Co content and the carrier density.

Metallic behavior in heavily doped semiconductors has been explained in general by the formation of a degenerate band of carriers provided by the dopants [38]. The metallic transition in our Co doped ZnO nanoparticles has been seen to be a combined effect of the Co ions and the oxygen vacancies, which act as donors and may lead to the formation of a shallow band of carriers [40,41]. We note that simulations and Density Functional Theory (DFT) calculations show that *uncharged* oxygen vacancies lead to deep level defects that may not hybridize with the d or p states. However it has also been shown [42] that *singly charged* oxygen vacancies ($V_O^+$) can promote ferromagnetism in ZnO by the overlap of the minority spin states of Co and the charged vacancy states. (A singly positively charged vacancy corresponds to one, rather than two electrons being trapped in the vicinity of the vacancy).

The suppression of this insulator-metal transition after annealing at higher oxygen pressure (see figure 6) clearly indicates that the carriers generated by the oxygen vacancies are the origin of the metallic behavior. However the absence of the MI transition even in the oxygen reduced pure ZnO and the systematic increase in its sharpness with increasing Co content, read alongside the lowering of the resistivity for increasing Co content (in the oxygen reduced samples), suggests that the metallicity occurs after the resistivity has decreased to a sufficiently low value. In other words metallicity may come about when the carrier concentration increases to a sufficiently high value. In a similar system, $TiO_2$, the critical concentration above which metallic behavior is observed is ~ $5 \times 10^{18}$ /cc [43]. We may therefore conclude that in our case the metallicity occurs at high temperatures for sufficiently high carrier concentrations caused by the combination of oxygen vacancies and Co ions.

A comparison of the data of resistivity as a function of temperature for the different samples shows that the large activation energy (0.3eV) is reduced by the forming gas to some extent (by about 0.06eV at most) but still remains quite high. The metallic transition cannot be explained as a thermally activated liberation of bound carriers with activation energy of 0.3eV into the conducting states, since the available thermal energy kT<<Ea. However a more realistic way of explaining the data appears to be to see the resistivity in the samples annealed in forming gas as comprising of two components; a semiconducting part with an activation energy close to 0.3 eV and a metallic part arising from the electrons trapped in the shallow defect states with binding energy comparable to 380K (~32meV), the temperature corresponding to the minimum of the resistivity. At low temperatures the semiconducting part is expected to be dominant since the defect states are still not emptied and secondly the semiconducting contribution to the resistivity is very large. As T approaches 380K from below, the defect states are emptied and the electrons spill over into the conduction band.  These defect states could be the *uncharged* oxygen vacancies $V_o$ that easily lose an electron to become a singly charged vacancy $V_o^{+1}$ [44]. The charged vacancy then can facilitate the FM interaction between the Co ions.

Does the metal-insulator transition shed any further light on the mechanism of the observed ferromagnetism? As discussed earlier, the resistivity of our samples is very large and remains in the insulating state for T<380K. If we consider the picture that the samples undergo a metal insulator transition for T~380K we see that the ferromagnetism (FM) at higher temperatures (T>380K) is coexisting with the metallic state where free carriers are present, while at lower temperatures the FM is present in the insulating state where such carriers, as available, are more or less localized. This is a scenario reminiscent of that proposed by Calderon *et. al* [45] for the FM in DMS, in particular Co doped $TiO_2$ and ZnO, that have insulating characteristics but with carriers weakly bound to defects. They envisage a scenario where the bound carriers mediate FM at low temperatures via the mechanism of overlap of bound magnetic polarons while at higher temperatures, $K_BT>E_B$, (where $E_B$ is the binding energy of the electron to the defect), the electrons are excited into the mobile, conducting states and may mediate the FM via a Zener-RKKY carrier mediated form. The carrier weakly bound to the oxygen vacancy, can be considered as the center of this distortion that polarizes the Co spins within its radius to form the spin polaron. The signal feature of this scenario is that if one relies only on the overlapping magnetic polarons as the source of FM and the weakly bound carriers as the source of the overlap between the Co- *d* states and the defect states, then at high enough temperatures where the defects will tend to release the trapped carriers, the system should move towards a *paramagnetic* metallic state with breakdown of the polaron picture. The presence of FM even within the metallic state at higher temperatures ($T_{MI}<T<T_c$) comes about in this picture [45] since the freed carriers can assume the role of the mediators of the RKKY type interaction between the Co ions. One of the attractive features of the theory is that it smoothly transposes between the low temperature bound magnetic polarons and higher temperature mobile, RKKY-Zener type mechanism. The effects of disorder on the RKKY interaction have further shown [46] that the phase diagram and $T_c$'s, obtained from the percolation picture in the context of the RKKY, assume realistic values when a cutoff is assumed in the RKKY interaction. Such a cutoff may arise from the effects of disorder or localization effects of carriers.

## IV. CONCLUSION

We have explored the combined effect of Co doping and oxygen vacancies in ZnO nanoparticles on the transport and optical properties in the compositions that have been established earlier to be free of metallic clusters. The optical studies further strengthen the conclusion that the Co ions are in substitutional sites while there are indications of defect states within the band gap. The band gap itself decreases systematically with Co content. While this effect may be attributed to the observed increase of the lattice constant with Co addition there may also be contributions coming from the enhanced sp-d overlap that has been discussed in the literature. A metal-insulator (MI) transition is observed only in the oxygen depleted Co doped ZnO nanoparticles. Two dominant features exist in the oxygen depleted Co doped ZnO nanoparticles. Firstly there is a remarkable decreased in the resistivity for all compositions in the oxygen reduced samples as compared to the samples without reduction. Secondly the resistivity systematically decreases with increasing Co content. In addition to the MI transition, the above significant and reversible characteristics are seen to be a combined effect of the Co ions and oxygen vacancies that we understand as leading to the formation of shallow donor states that may hybridize strongly with the Co *d* states to stabilize the RTFM. The relevance of a theoretical framework that supports the transition from ferromagnetism via overlapping of spin polarons at lower temperatures in the insulating state, to a more itinerant electron mediation of ferromagnetism at higher temperatures has been discussed in the context of the observed metal insulator transition. Experiments are underway to explore the resistive behaviour in thin film samples of the Co doped compositions and in Cu and Al co-doped materials with variations in the electron concentration.

**ACKNOWLEDGMENTS**
SKH acknowledges support to this work from the project *Development and Study of Magnetic Nanostructures*, Higher Education Commission, Government of Pakistan. MN acknowledges financial support from the Pakistan Science Foundation Grant No: NSFC/RES/Phys(18). The authors are grateful to Prof. S. Ismat Shah, Department of Physics and Astronomy, University of Delaware, Newark, Delaware 19716, USA. for providing the transmission electron microscopy (TEM) data.

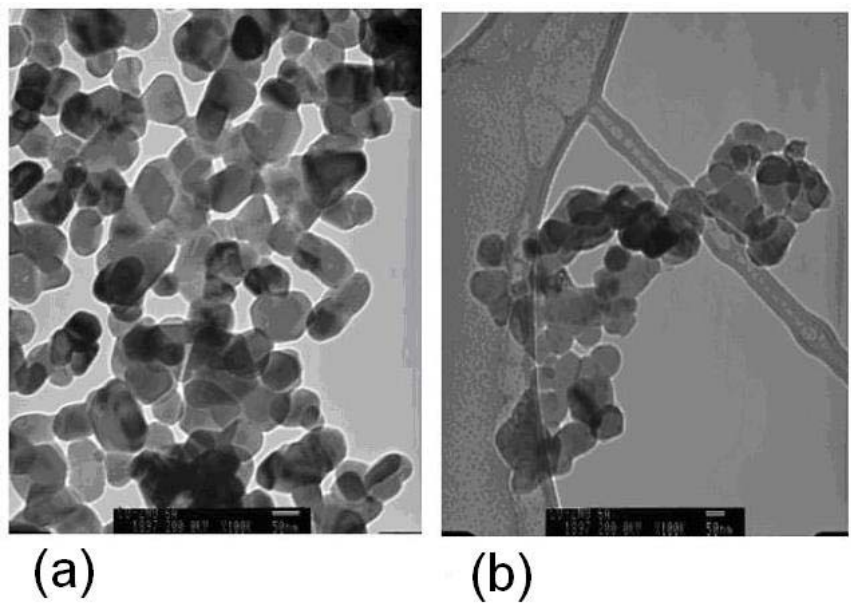

Figure 1: TEM images of Zn$_{0.94}$Co$_{0.06}$O nanoparticles annealed in (a) forming gas; (b) Air.

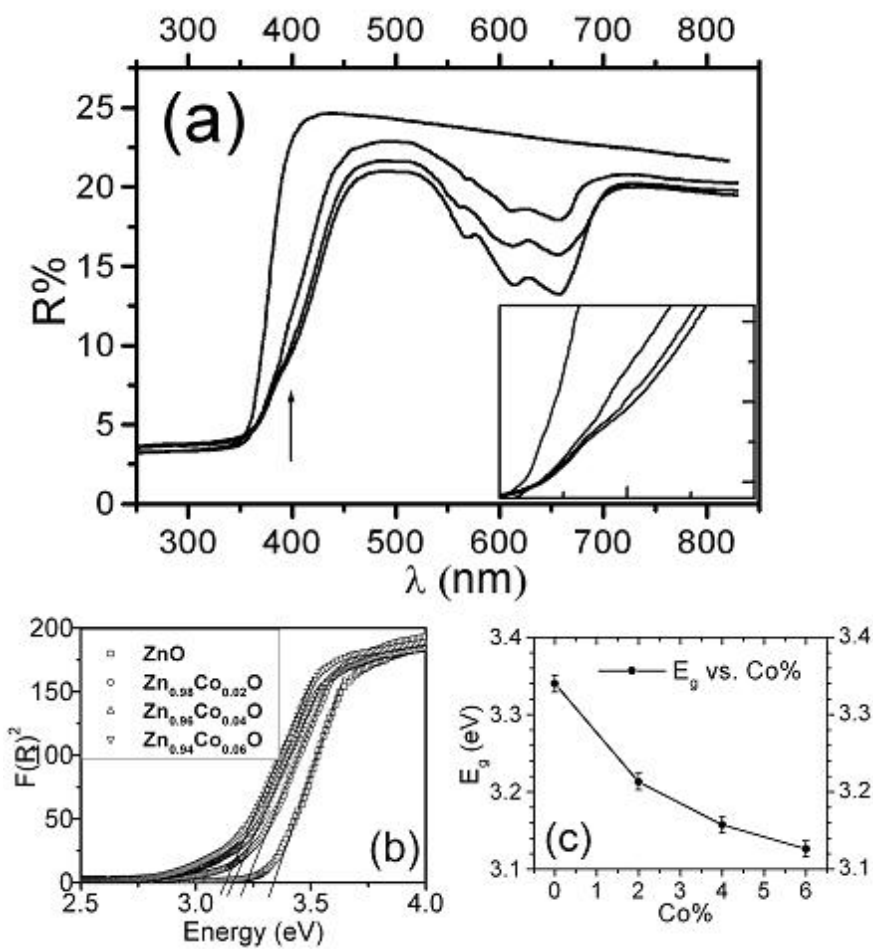

Figure 2: (a) Room temperature optical reflectance spectra of $Zn_{1-x}Co_xO$ (x=0.00 to x=0.06) nanoparticles. Concentrations are: x = 0.00 top curve; x = 0.02 second from top; x = 0.04 third from top; x = 0.06 bottom curve. The observed minima in Co doped concentrations for $E<E_g$ are discussed in text. Arrow indicates the region shown expanded in the inset. Inset: Expanded view of the structure observed in the reflectivity for energies close to the band edge (see text for detail); (b) Plot of $F(R)^2$ vs $E$ (see text for details); (2c) Plot of band gap $E_g$ obtained from figure 2a vs Co concentration.

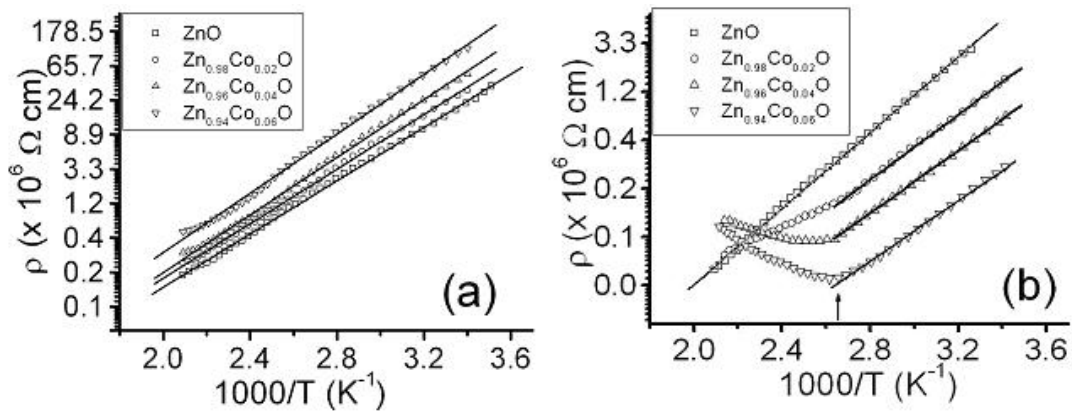

Figure 3: Temperature dependence of the resistivity of Co doped ZnO nanoparticles with varying Co concentrations (a) Annealed in air (b) Annealed in forming gas. Arrow (in figure b) indicates the point of metal-insulator transition (see text for details). Straight lines are fit to the Arrhenius equation.

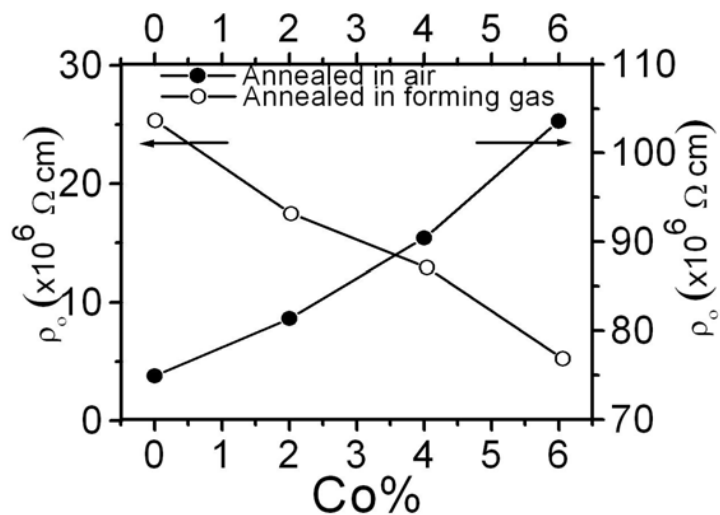

Figure 4: Resistivity prefactor $\rho_o$ ($\rho = \rho_o \exp(E_a/k_B T)$) versus Co concentrations for samples annealed in air (filled circles) and those annealed in forming gas (open circles), from the resistivity data data of figure 3.

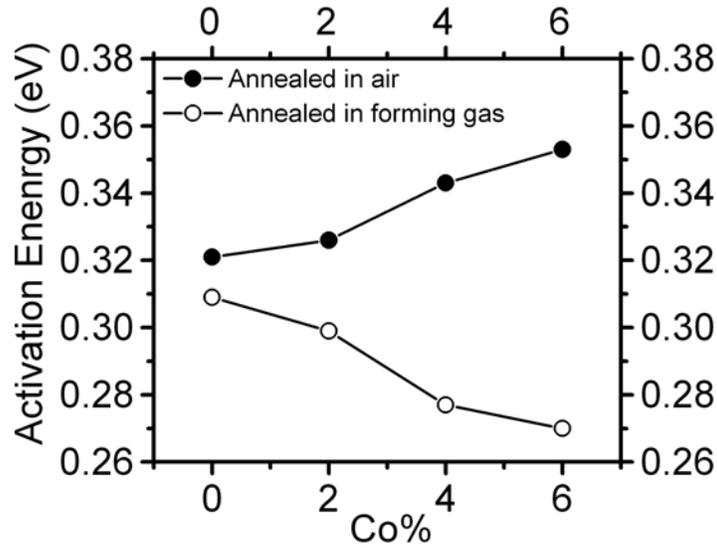

Figure 5: Activation energy (obtained form the fit to the Arrhenius equation) versus Co concentration for samples annealed in air (filled circles) and those annealed in forming gas (open circles). The differing trends in the two cases are evident.

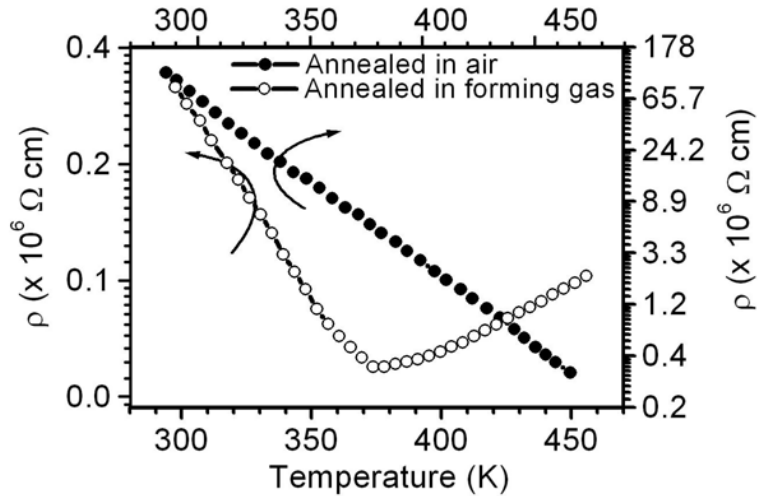

Figure 6 Resistivity versus temperature of the 6% Co doped sample showing the reversible effects of forming gas anneal (open circles) and subsequent oxygen anneal (filled circles). The MI transition is quenched by the oxygen anneal while the resistivity increases many folds.